\documentclass[conference]{IEEEtran}
\usepackage{cite}
\usepackage{amsmath,amssymb,amsfonts}
\usepackage{algorithmic}
\usepackage{graphicx}
\usepackage{textcomp}
\usepackage{xcolor}
\usepackage{rotating}
\usepackage{multirow}    % For multi-row cells
\usepackage[margin=1in]{geometry}
\usepackage{hyperref}
\usepackage[sort, numbers]{natbib}

\def\BibTeX{{\rm B\kern-.05em{\sc i\kern-.025em b}\kern-.08em
    T\kern-.1667em\lower.7ex\hbox{E}\kern-.125emX}}

\begin{document}

\title{Agile Management for Machine Learning: A Systematic Mapping Study}

% \and
\author{
Lucas Romao\textsuperscript{1}, Hugo Villamizar\textsuperscript{2}, Romeu Oliveira\textsuperscript{1},
Silvio Alonso\textsuperscript{1}, Marcos Kalinowski\textsuperscript{1}\\
\textsuperscript{1}ExACTa Software Engineering Laboratory, Department of Informatics\\
Pontifical Catholic University of Rio de Janeiro (PUC-Rio), Rio de Janeiro, Brazil\\
\textsuperscript{2}fortiss GmbH, Munich, Germany\\
Emails: \{lromao, rferreira, smarques, kalinowski\}@inf.puc-rio.br, guarinvillamizar@fortiss.org
}

% romeu@tecgraf.puc-rio.br

\maketitle

\begin{abstract}

[Context] Machine learning (ML)-enabled systems are present in our society, driving significant digital transformations. The dynamic nature of ML development, characterized by experimental cycles and rapid changes in data, poses challenges to traditional project management. Agile methods, with their flexibility and incremental delivery, seem well-suited to address this dynamism. However, it is unclear how to effectively apply these methods in the context of ML-enabled systems, where challenges require tailored approaches.
[Goal] Our goal is to outline the state of the art in agile management for ML-enabled systems.
[Method] We conducted a systematic mapping study using a hybrid search strategy that combines database searches with backward and forward snowballing iterations. 
[Results] Our study identified 27 papers published between 2008 and 2024. From these, we identified eight frameworks and categorized recommendations and practices into eight key themes, such as Iteration Flexibility, Innovative ML-specific Artifacts, and the Minimal Viable Model. The main challenge identified across studies was accurate effort estimation for ML-related tasks.
[Conclusion] This study contributes by mapping the state of the art and identifying open gaps in the field. While relevant work exists, more robust empirical evaluation is still needed to validate these contributions.
% such as data management and sequential dependencies in workflows—

\end{abstract}

\begin{IEEEkeywords}
agile management, machine learning, artificial intelligence, systematic mapping study
\end{IEEEkeywords}

\section{Introduction}

% Context

Machine Learning (ML), a subset of Artificial Intelligence (AI), involves building systems that learn from data using statistical methods, evolving through iterative experimentation and adaptation \cite{mitchell1997machine}. In recent years, ML has been widely integrated into software of all types. This growth highlights ML's dynamic nature, which demands flexible development processes to manage uncertainty and rapid changes \cite{nahar2023meta-summary}. Agile methodologies, such as Scrum \cite{schwaber1997scrum} and Kanban \cite{ahmad2013kanbansms}, conceptually align with the iterative, exploratory nature of ML workflows, especially during data preparation and model tuning.

However, applying Agile to ML poses challenges. Time-boxed sprints struggle to accommodate ML's unpredictable experimentation, while sequential dependencies in ML workflows conflict with Agile's preference for small independent increments \cite{nahar2023meta-summary, cohn2004userstoriesapplied}. Bridging this gap requires tailored adaptations of Agile practices to address ML-specific complexities, such as evolving data requirements and model uncertainty. Despite growing interest in this area, there remains a lack of consolidated knowledge about how the integration of Agile and ML practices is being approached, what challenges persist, and which practices are proving effective.

% Goal and Methods
To address this need, this paper presents a systematic mapping study (SMS) to synthesize existing research on agile management for ML-enabled systems. Our objective is to characterize the state of the art, by identifying proposed approaches, practices and adaptations, recommendations, challenges, and research gaps. 

% Results
The SMS identified 27 peer-reviewed papers published between 2008 and 2024. From this corpus, we identified eight approaches designed to adapt agile methodologies for ML-enabled systems development processes. Furthermore, by applying thematic synthesis \cite{cruzes2011recommended}, we categorized practices and adaptations and recommendations into eight key themes.

First, \textit{Iteration Flexibility} enables data processing and model refinement through flexible sprints. Second, \textit{ML-Specific Artifacts}, such as \textit{Model/Data Stories} \cite{A02-vaidhyanathan2022agile4mls} and \textit{Ethical User Stories} \cite{A09-kemell2022utilizing}, bridge gaps between traditional agile documentation and ML requirements. Third, \textit{Decoupled Ceremonies} detach Scrum rituals schedule from flexible sprint's pace to allow equally spaced feedback. Fourth, \textit{Hybrid Approaches} harmonize data mining workflows with agile rhythms to reconcile sequential dependencies and iterative delivery. Fifth, \textit{MVM or Demo API} delivers (minimal viable) functional models for hypothesis testing and for software teams, preserving agile deliverables amid ongoing experimentation. Sixth, with \textit{Kanban Adoption} ML teams prioritize Kanban over Scrum as their primary agile framework. Seventh, \textit{Business Alignment} ensures collaboration between technical teams and stakeholders to focus on value-adding tasks. Finally, \textit{Ethical Considerations} integrate guidelines for backlog management.

Furthermore, we synthesized three distinct challenge themes. First, \textit{Sprint Planning \& Effort Estimation}. Second, \textit{Methodological, Training, and Strategic Alignment Deficiencies}. Finally, \textit{Ethical Considerations}, with the foremost being effort estimation for ML-related tasks.

The remainder of this paper is organized as follows: Section II provides background on Agile methodologies and challenges in ML-enabled systems. Section III details the SMS protocol, including research questions and search strategy. Sections IV–V present results and discussions, respectively. Section VI addresses threats to validity, and Section VII concludes with implications for research and practice.

\section{Background and Related Work}

Driven by the need for flexibility and faster iterative delivery of value, a group of software practitioners introduced \textit{The Agile Manifesto} \cite{beck2001manifesto}, which became a foundation for agile methods, one of the most widely used approaches in modern software development \cite{whatMakesAgileSoft.Agile}. As software systems have become increasingly complex and data-driven, agile has been extended to new domains, including ML-enabled systems. Agile methods seem well suited to address the dynamic nature of ML development, characterized by experimental cycles and rapid changes in data. However, despite the widespread adoption of agile methods, its application in managing ML-enabled projects brings challenges, including obstacles to process and people management in ML projects \cite{nahar2023meta-summary}. 

On the process side, the experimental and uncertain nature of ML development complicates effort estimation, making it difficult to guarantee deliverables within sprints. Unpredictable results and variables can derail planned work, while the lack of coherence between ML-enabled and non-ML components often leads to backlog disorganization and development delays. On the people side, gaps in mutual understanding pose significant barriers. Data Science (DS) teams often lack familiarity with Software Engineering (SE) principles, leading to suboptimal coding practices, while developers may struggle to grasp ML concepts, resulting in misaligned expectations and communication breakdowns \cite{busquim2024interaction}.

These challenges underscore the need to better understand how agile project management can be tailored to ML-enabled systems. Although we identified related work concerning a multivocal literature review exploring practices for managing ML products in a more general sense with a focus on operationalizing ML in production environments \cite{alves2023practices}, our mapping study focuses specifically on agile project management and on how agile methods are adapted to address the unique challenges of ML-enabled system development.

\section{Systematic Mapping Protocol}

In this section, we present the research goal, questions, search strategy, and inclusion and exclusion criteria of our SMS, which follows the guidelines proposed by Peterson \textit{et al.} \cite{petersen2015guidelines}, widely used in SE research.

\subsection{Research Objectives and Questions}

The main objective of this SMS is \textbf{to outline the state of the art in Agile Management for ML-enabled Systems}. Based on this goal, we propose the following Research Questions (RQ).

\textbf{RQ1. What agile management approaches have been proposed for ML-enabled systems development?} This question aims to uncover well-established agile management approaches used during ML-enabled systems development.

\textbf{RQ2. How are agile management practices tailored to align with ML-enabled systems development?}  This question aims to uncover how well-established agile management practices were modified or created better to fit the demands of ML-enabled systems development.

\textbf{RQ3. What recommendations have been proposed for managing ML-enabled systems development?} This question aims to identify and summarize the various recommendations in the literature for managing the development of ML-enabled systems. 

\textbf{RQ4. What are the reported challenges when adapting agile management practices to ML-enabled systems development?}  This question aims to identify 
open challenges in Agile Management for ML-enabled systems development.

\textbf{RQ5. How can the research contributions be classified?} This question aims to classify the research type according to the classification scheme proposed by Wieringa \textit{et al.} \cite{wieringa2006requirements}.

\textbf{RQ6. What empirical methods have been applied to validate the research contributions?} This question aims to investigate the empirical methods that have been carried out. Gathering this information provides an initial overview of the applied scientific rigor of the identified studies.

%\cut{The first four questions aim to outline the conceptual aspects of the publications, while the last two focus on evaluating the scientific nature of the presented evidence.}

\subsection{Search Strategy}

In this SMS, we adopted the search strategy proposed by Wohlin \textit{et al.} \cite{wohlin2022successful}, which has been recommended as effective for identifying relevant primary studies. It combines a structured search in the Scopus\footnote{\href{https://www.scopus.com/home.uri}{https://www.scopus.com/home.uri}} digital database using a specific search string, supplemented by iterative backward and forward snowballing. Forward snowballing involves identifying studies that have cited a particular paper, allowing researchers to find more recent related research. Backward snowballing, on the other hand, examines the references within a study to locate earlier relevant research. 

To perform the initial database search on Scopus, we formulated the search string following the PICO (Population, Intervention, Comparison, Outcomes) strategy \cite{leonardo2018pico}. In our study, ML-enabled Systems constitute the \textit{population}, and we seek to examine Agile Management practices for such systems (\textit{intervention}). Since this study does not involve comparing studies, an outcome variable for comparison was not required. Based on this strategy, the final search string was: \textit{(``machine learning" OR ``artificial intelligence") AND ((``management" OR ``practices") AND (``agile" or ``scrum"))}.

\subsection{Study Selection}

We set the timeframe for our search to the end of 2024. Aiming to identify evidence that meets our research goal, the Inclusion Criteria (IC) for our study is to include papers that address agile management practices and approaches used in ML-enabled systems development. In cases where multiple papers report on the same study, we considered only the most recent one. The Exclusion Criteria (EC) consider excluding papers from our IC to maintain a consistent quality standard for the selected papers. The EC applied in this study is outlined in Table \ref{icec-criteria}. Details of the IC and EC application process are available in our supplementary material, available in our Zenodo Open Science  Repository\footnote{\url{https://doi.org/10.5281/zenodo.14105692}}.

\begin{table}[h!]
\centering
\scriptsize
\caption{EXCLUSION CRITERIA}
\label{icec-criteria}
\begin{tabular}{|p{1cm}|p{6cm}|}
\hline
\multicolumn{1}{|c|}{\textbf{Criteria}} & \multicolumn{1}{c|}{\textbf{Description}} \\ \hline
EC1 & Papers that do not meet the Inclusion Criteria. \\ \hline
EC2 & Papers that address ML/AI techniques for Agile Management, given that, is not what we're looking for. \\ \hline
EC3 & Papers that do not address ML-enabled systems development (e.g., IoT projects, game development, and other software development).\\ \hline
EC4 & Papers not written in English. \\ \hline
EC5 & Conference reviews, posters, abstracts, short papers (less than four pages), and presentations. \\ \hline
EC6 & Grey literature such as blogs, theses, and papers not reviewed by peers. \\ \hline
\end{tabular}
\end{table}

\begin{figure}[h!]
    \centering
    \includegraphics[width=18.5pc]{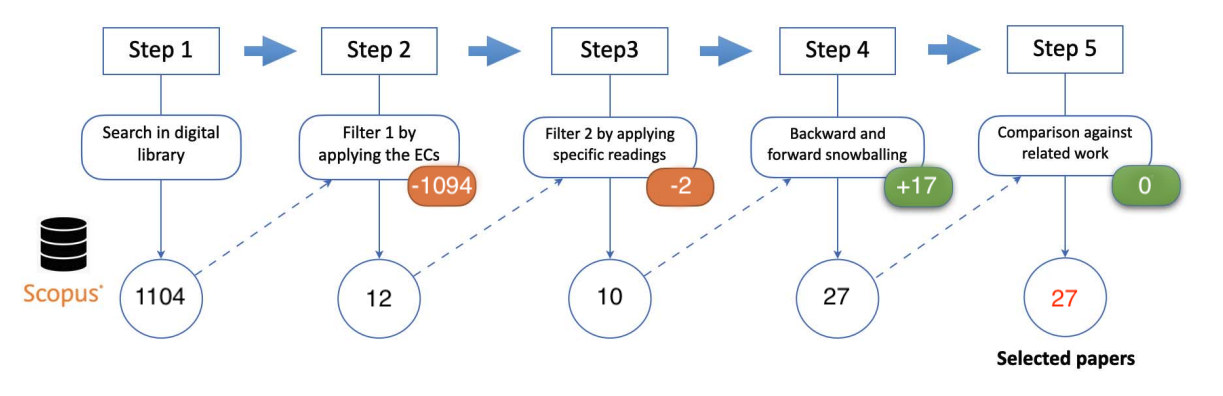}
    \caption{Paper selection process}
    \label{searchprocess}
\end{figure}

\begin{figure}[h!]
    \centering
    \includegraphics[width=18.5pc]{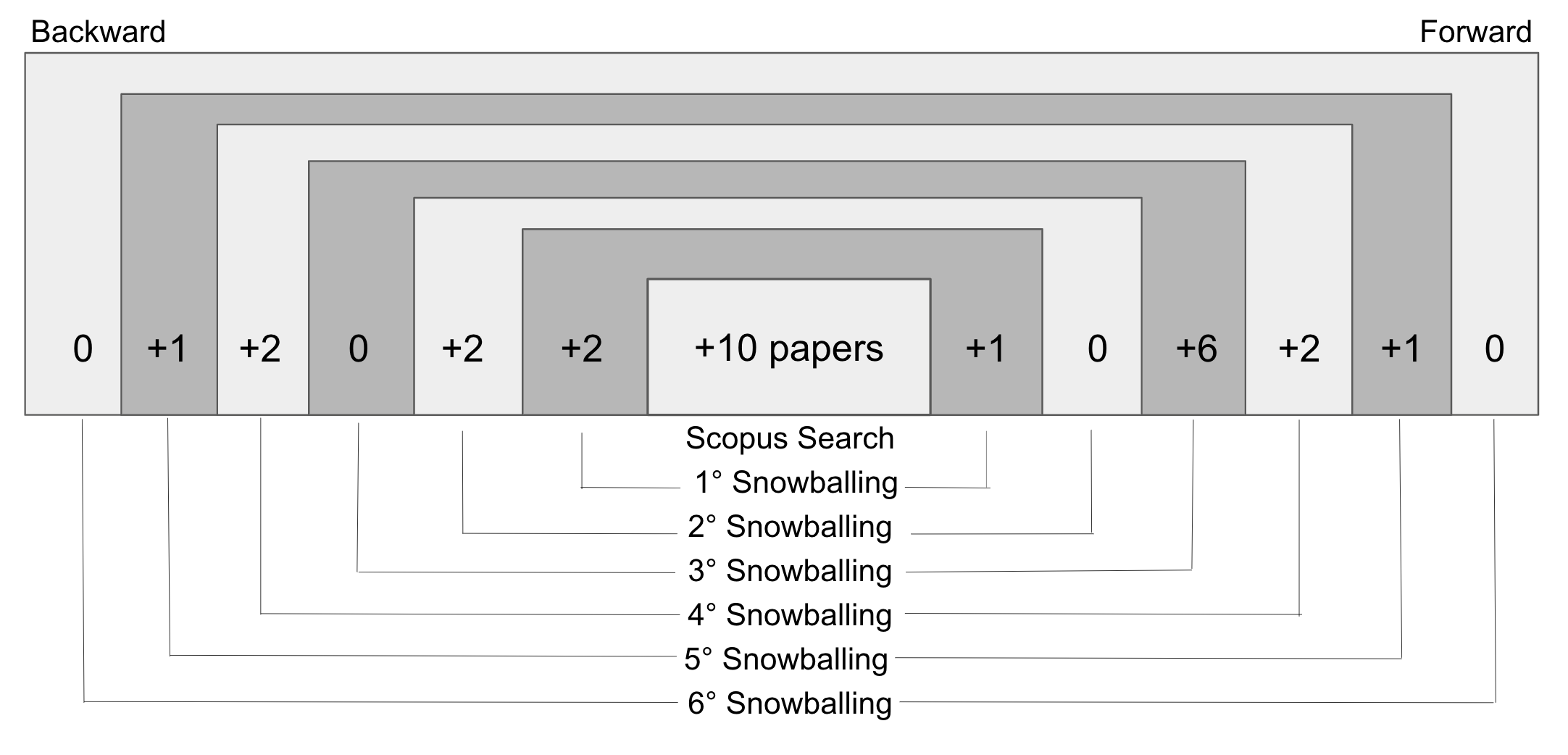}
    \caption{Iterative snowballing}
    \label{snowballing}
\end{figure}

The selection process, as illustrated in Fig. \ref{searchprocess}, is detailed hereafter. First, we searched for papers using the defined search string in the Scopus Library, filtering by titles, abstracts, and keywords. Of the 1,104 papers initially retrieved, only ten met the IC. This process was conducted by the main author and subsequently reviewed by the other authors.

Secondly, we applied backward and forward snowballing (via Google Scholar) on each selected paper to identify additional relevant studies not captured by the initial Scopus search. This iterative process is illustrated in Figure \ref{snowballing}. Through snowballing, we identified 17 additional papers, bringing the total number of selected studies to 27. In total, we screened over 2,400 papers across the Scopus search and snowballing iterations.

\subsection{Data Extraction and Classification Scheme}

The Data Extraction and Classification Scheme for each paper is outlined in Table \ref{data-form}. The selection process and the extracted data are documented in our online Zenodo repository, which includes information on each identified paper, the reason for its inclusion or exclusion, and the data extracted to answer each RQ. 

\begin{table}[h]
\centering
\scriptsize
\caption{DATA EXTRACTION FORM}
\label{data-form}
\begin{tabular}{|p{=1.5cm}|p{5cm}|}
\hline
\multicolumn{1}{|c|}{\textbf{Information}}  & \multicolumn{1}{c|}{\textbf{Description}} \\ \hline
Study metadata & Includes the title of the paper and information such as venue, type of venue, and year of publication. \\ \hline
Agile approaches (RQ1) & Approaches and frameworks created to suit ML-enabled system management demands. \\ \hline
Approaches' practices (RQ2) & Practices created or adapted to suit ML-enabled systems management demands, such as new concepts of the sprint or new artifacts to manage the backlog. \\ \hline
Reco\-mmen\-dation (RQ3) & Recommendations from papers that do not propose a new approach but provide insights on how to manage ML-enabled system development using agile methodologies. \\ \hline
Challenges (RQ4) & Challenges that arise when adopting agile management methodologies in ML-enabled system development. \\ \hline
Research type facet (RQ5) &  Classification of paper's research type proposed by Wieringa \textit{et al.} \cite{wieringa2006requirements}: evaluation research, solution proposal, philosophical paper, opinion paper, or experience paper. \\ \hline
Empirical Evaluation (RQ6) & Classification of the empirical strategy proposed by Wohlin \textit{et al.} \cite{wohlin2024experimentation}: experiment, case study, survey. \\ \hline
\end{tabular}
\end{table}

\section{Results}

\subsection{RQ1. What agile management approaches have been proposed for developing ML-enabled systems?}

We identified eight agile management approaches to deal with ML-enabled systems development. We summarized them in Table \ref{rq1&2}.

\begin{table}[h]
\caption{Agile Management Approaches and Their Adaptations}
\centering
\begin{tabular}{|p{=1.5cm}|p{5cm}|}
\hline
\textbf{Approach} & \textbf{Adaptations} \\ \hline

\multirow{7}{=}{Agile4MLS \cite{A02-vaidhyanathan2022agile4mls}} 
    & Model \& Data Stories \\ \cline{2-2}
    & Demo ML-API \\ \cline{2-2}
    & One-week Sprints \\ \cline{2-2}
    & Differentiation between ML \& non-ML Intensive Product Backlog Items (PBI). \\ \cline{2-2}
    & ML Team Dailies with a rotating software (SW) team member, and vice-versa. \\ \cline{2-2}
    & Separate ML and SW teams working in parallel, with ML PBIs planned 2 sprints ahead cover dependencies. \\ \cline{2-2}
    & One Weekly with ML \& SW Teams together \\ \hline

\multirow{3}{=}{STAMP 4 NLP \cite{A03-kohl2021stamp}} 
    & Development Loop (hours-days) to handle granular model development \\ \cline{2-2}
    & Evolution Loop (two-four weeks) to handle larger PBIs and CI/CD \\ \cline{2-2}
    & BizDev \& CI/CD alignment \\ \hline

\multirow{6}{=}{SKI \cite{A11-saltz2019ski}} 
    & Capability-based Iterations \\ \cline{2-2}
    & Decoupled Scrum Ceremonies \\ \cline{2-2}
    & WIP limits \\ \cline{2-2}
    & SKI Master role to manage flow and WIP limits \\ \cline{2-2}
    & Testable Hypothesis alongside each US \\ \hline

\multirow{4}{=}{Scrum-DS \cite{A12-baijens2020applying}} 
    & Sprint 0 for Data Preparation \\ \cline{2-2}
    & One CRISP-DM full cycle per Sprint (Hybrid-Approach Example) \\ \cline{2-2}
    & Four-week Sprint \\ \cline{2-2}
    & Demo ML-API \\ \hline

\multirow{4}{=}{Data Driven Scrum \cite{A18-saltz2022achieving}} 
    & Capability-based Iterations \\ \cline{2-2}
    & Decoupled Scrum Ceremonies \\ \cline{2-2}
    & T-Shirt sizing PBI Estimation \\ \cline{2-2}
    & Vertical/Horizontal Slicing \\ \hline

\multirow{2}{=}{Agile-facilitated KD \cite{A22-schmidt2018synthesizing}} 
    & One CRISP-DM full cycle per Sprint (Hybrid-Approach Example) \\ \cline{2-2}
    & An improved model version after each Sprint \\ \hline

\multirow{2}{=}{ADS \cite{A23-lei2020agile}} 
    & Scrumban + CRISP-DM cycle \\ \cline{2-2}
    & Every phase in the process is rollbackable \\ \hline

\multirow{3}{=}{ASD-DM \cite{A24-alnoukari2008applying}} 
    & Speculate-Collaborate-Learn cycle instead Build-Measure Learn \\ \cline{2-2}
    & Rollback CRISP-DM \\ \cline{2-2}
    & Short Iterations (one-two weeks) \\ \hline

\multirow{2}{=}{None \cite{A09-kemell2022utilizing}} 
    & Adapt User Stories (US) to address ethical concerns \\ \cline{2-2}
    & Extend US template to: ``\textit{AS A, I WANT TO, IN ORDER, GIVEN} [ethical context]'' \\ \hline

\end{tabular}
\label{rq1&2}
\end{table}

\subsection{RQ2. How are agile management practices tailored to align with the development of ML-enabled systems?}

Each management approach identified in RQ1 tailored some agile practices to fit the needs of ML-enabled systems. The authors have presented adaptations such as new ways to manage backlogs, propose an integration of CRISP-DM and Scrum Framework, and, in some cases, decoupled Scrum ceremonies from sprints. The adaptations are also reflected in Table \ref{rq1&2}.

\subsection{RQ3. What recommendations have been proposed for managing ML-enabled systems development?}

Given the variety of identified approaches, adaptations, and recommendations, we applied thematic synthesis with open coding and summarized the codes into key themes \cite{cruzes2011recommended}. This process was carried out by the first author and reviewed by the last author. It is transparently documented in our open science repository. As a result, eight key themes emerged, which we describe hereafter.

First, \textbf{Iteration Flexibility} \cite{A03-kohl2021stamp, A10-saltz2017comparing, A11-saltz2019ski, A14-kraut2022application, A18-saltz2022achieving, A19-saltz2021identifying, A26-10.1007/978-3-031-48550-3_6} manifests through capability-based iterations or flexible sprints, addressing the inherent unpredictability of ML experimentation cycles. This adaptation suits extended data processing phases and model refinement tasks that fail to fit in conventional time-boxing. 

Second, the \textbf{ML-Specific Artifacts} \cite{A02-vaidhyanathan2022agile4mls, A06-singla2018analysis, A07-jackson2019agile, A09-kemell2022utilizing}, such as Model Stories, Data Stories, and Ethical User Stories, introduce new requirements engineering tools aiming at bridging the gap between traditional agile documentation and ML needs. 

Third, \textbf{Decoupled Ceremonies} \cite{A11-saltz2019ski, A18-saltz2022achieving, A19-saltz2021identifying} reconfigure scrum by detaching well-known rituals from fixed sprint cycles. This structural shift accommodates ML's development pace, enabling teams to maintain continuous feedback without forcing artificial synchronization with incomplete experimental outcomes. 

Fourth, on \textbf{Hybrid Approaches} \cite{A01-vial2023managing, A08-haakman2021ai, A12-baijens2020applying, A13-qadadeh2020improved, A20saltz2020identifying, A22-schmidt2018synthesizing, A23-lei2020agile, A24-alnoukari2008applying, A25-uysal2022machine, A27uysal2023toward}, some approaches propose an alignment between data mining workflows and agile methods, these integrations help reconcile ML's sequential dependencies with agile's iterative delivery expectations. 

Fifth, \textbf{MVM or Demo API} \cite{A02-vaidhyanathan2022agile4mls, A07-jackson2019agile, A12-baijens2020applying, A23-lei2020agile} contributes to the development process by providing small, but functional models that can be used by the SW team or, to test a business hypothesis. These models help to ensure a traditional agile deliverable concept while the ML team continues on the experimentation.

Sixth, \textbf{Kanban adoption} \cite{A10-saltz2017comparing, A16-baijens2020data, A17-saltz2019achieving, A25-uysal2022machine} points out that the usage of Kanban could be more efficient than Scrum due to its absence of Sprints, which is one of the most challenging aspects to adapt for ML development.

Seventh, \textbf{Business alignment} \cite{A13-qadadeh2020improved, A15-dastgerdi2021appropriate, A21-lahiri2023evaluating} aims at ensuring that Data Scientists and ML Engineers are aligned with business needs. By addressing this theme, we avoid unnecessary experimentation tasks and uncertainty-solving problems that will not add value for stakeholders.

Finally, \textbf{Ethical Considerations} \cite{A04-halme2021write, A05-leijnen2020agile, A09-kemell2022utilizing} are important in preventing errors that may result in undesirable model behavior.

% Hybrid approaches  —  10
% Iteration Flexibility  —  7
% MVM as Demo API  —  5
% Innovative ML artifacts  —  4
% Kanban adoption  —  4
% Ethical Considerations  —  3
% Decoupled ceremonies—3
% BizML alignment  —  3

\subsection{RQ4. What are the reported challenges when adapting agile management practices for developing ML-enabled systems?}

The majority of papers presented challenges in managing ML-enabled systems with agile management practices. As in RQ3, we applied thematic synthesis with open coding and summarized the challenges into three key themes.

\textbf{Sprint planning} \cite{A15-dastgerdi2021appropriate, A21-lahiri2023evaluating, A22-schmidt2018synthesizing, A23-lei2020agile}  \& \textbf{Effort estimation} \cite{A06-singla2018analysis, A10-saltz2017comparing, A11-saltz2019ski, A12-baijens2020applying, A14-kraut2022application, A17-saltz2019achieving, A19-saltz2021identifying, A21-lahiri2023evaluating, A25-uysal2022machine, A26-10.1007/978-3-031-48550-3_6}. Multiple studies highlight difficulties in accurately estimating effort due to the experimental nature of ML workflows, leading to misalignments between data preparation and modeling activities and sprint planning ceremonies. Short sprint cycles (e.g., two weeks) struggle to accommodate time-consuming ML-related activities. Researchers \cite{A12-baijens2020applying, A19-saltz2021identifying, A22-schmidt2018synthesizing} observed that experimental ML tasks often exceed sprint boundaries, and rigid time-boxing of data preparation phases disrupted workflow continuity. Furthermore, the inherently sequential and interdependent nature of ML workflows (\textit{e.g.}, data preprocessing, model training, model evaluation) contrasts with the discrete, independent structure of traditional User Stories \cite{cohn2004userstoriesapplied}, leading to mismatches in sprint planning. These factors contribute to ML tasks recurring across multiple sprints.

\textbf{Methodological} \cite{A01-vial2023managing, A03-kohl2021stamp, A16-baijens2020data, A18-saltz2022achieving, A20saltz2020identifying, A21-lahiri2023evaluating, A23-lei2020agile, A25-uysal2022machine, A27uysal2023toward}, \textbf{Training} \cite{A04-halme2021write, A08-haakman2021ai, A26-10.1007/978-3-031-48550-3_6}, and \textbf{Strategic Alignment Deficiencies} \cite{A11-saltz2019ski, A13-qadadeh2020improved, A19-saltz2021identifying}. Key challenges include the lack of generalized frameworks for ML-specific agile practices and practitioners tendency to selectively adopt ("cherry-pick") practices without structured guidance. Scholars \cite{A02-vaidhyanathan2022agile4mls, A15-dastgerdi2021appropriate, A18-saltz2022achieving, A21-lahiri2023evaluating} further identify empirical limitations such as reliance on single case studies and limited confidence in the generalizability of proposed solutions. Additional challenges stem from misalignment between business stakeholders, often misunderstanding ML capabilities, while data scientists face difficulties translating technical constraints into actionable business requirements. This underscores the necessity of alignment between business objectives and DS activities. Moreover, effective collaboration requires cross-training initiatives, where both ML and SE teams develop mutual competencies in DS and SE practices~\cite{busquim2024interaction}.

\textbf{Ethical Considerations} \cite{A04-halme2021write, A09-kemell2022utilizing, A21-lahiri2023evaluating}. While AI ethics are researched, practical implementation strategies are lacking in Agile frameworks, potentially leaving ethical aspects underexplored during agile ML-enabled system development.

\subsection{RQ5. How can the research contributions be classified?}

\begin{figure}[h!]
    \centering
    \includegraphics[width=18.5pc]{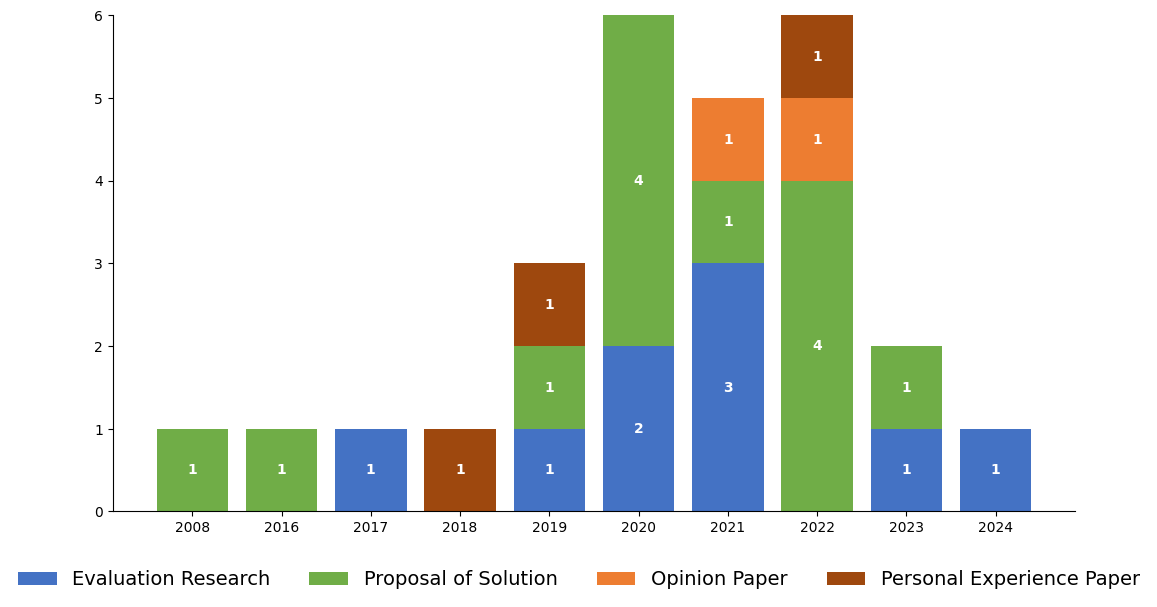}
    \caption{Distribution of research type per year}
    \label{researchperyear}
\end{figure}

Fig. \ref{researchperyear} illustrates the annual distribution of research types across the reviewed literature following the classification framework by Wieringa \textit{et al.} \cite{wieringa2006requirements}. The analysis reveals that \textit{Solution Proposals} constitute the predominant research type (13 of 27 papers), followed by \textit{Evaluation Research}, accounting for nine papers. \textit{Personal Experience Papers} and \textit{Opinion Papers} together represent five papers. The subsequent section addresses the empirical evaluation methods used in these studies.

\subsection{RQ6. What empirical methods have been applied to validate the research contributions?} 

\begin{figure}[h!]
    \centering
    \includegraphics[width=18.5pc]{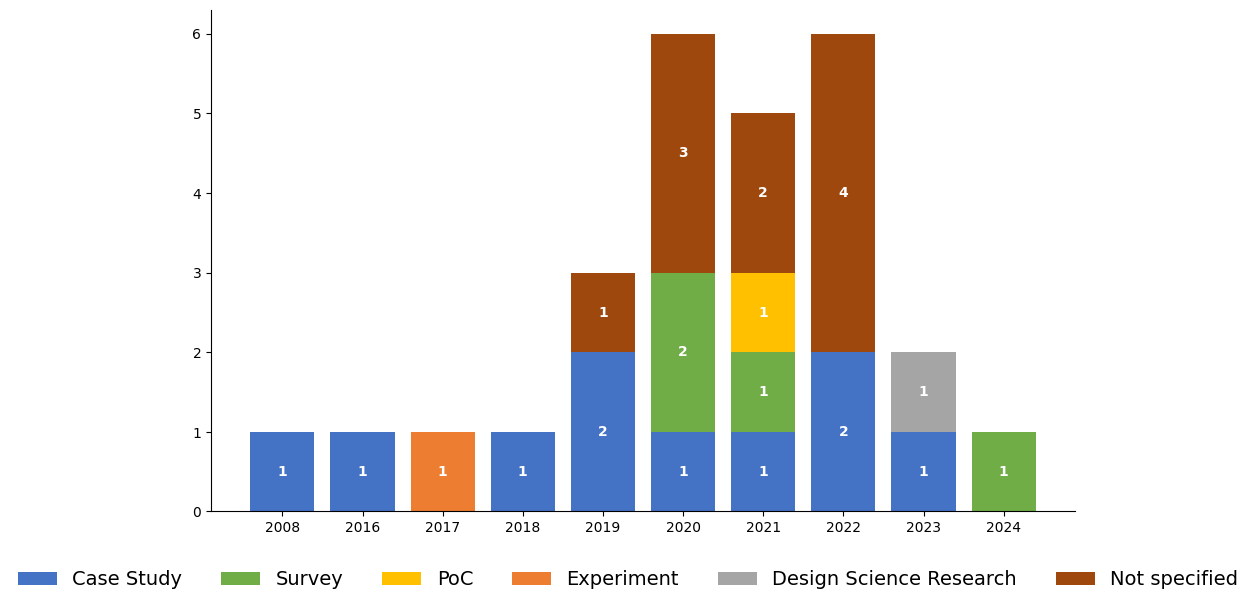}
    \caption{Distribution of evaluation type per year}
    \label{evaluationperyear}
\end{figure}

Fig. \ref{evaluationperyear} delineates the distribution of empirical evaluation methods across the analyzed studies. Of the 27 papers, 17 incorporate empirical evaluations, with \textit{Case Studies} constituting the majority (9 papers), followed by \textit{Surveys} (5 papers) and a single \textit{Experiment}. This underscores a scarcity of more rigorous empirical methods within the corpus, a concern given their critical role in validating discoveries. The corpus includes one \textit{Proof-of-Concept} study, which serves to demonstrate the feasibility of a proposed method, and one \textit{Design Science Research} study, which serves to develop and evaluate innovative artifacts aimed at addressing specific practical challenges. Conversely, a significant proportion of studies (9 of 27) lack empirical evaluation entirely; these consist predominantly of \textit{Personal Experience Papers}, \textit{Opinion Papers}, or \textit{Solution Proposals} without supporting evidence.

\section{Discussion}

Several agile management practices and approaches tailored for ML-enabled systems have been proposed, mostly in recent years. The contributions reflect both innovative adaptations and significant challenges that persist in operationalizing the unique demands of ML development. The literature encompasses a diverse set of agile management approaches (RQ1 \& RQ2) including Agile4MLS \cite{A02-vaidhyanathan2022agile4mls}, STAMP 4 NLP \cite{A03-kohl2021stamp}, SKI \cite{A11-saltz2019ski}, Scrum-DS \cite{A12-baijens2020applying}, Data Driven Scrum \cite{A18-saltz2022achieving}, among others (\cite{A18-saltz2022achieving, A22-schmidt2018synthesizing, A23-lei2020agile, A24-alnoukari2008applying}). 

RQ3 presents identified recommendations clustered by categories. Notably, \textit{Hybrid Agile-Lifecycle Integration} emerged as the most prominent recommendation type (10 of 27 papers), reflecting efforts to reconcile iterative agile practices with the sequential nature of ML workflows. This aligns with prior observations about the incompatibility of rigid agile frameworks with ML experimentation cycles \cite{A12-baijens2020applying, A22-schmidt2018synthesizing}. 

While these integrations address ML workflow sequentially, their success hinges on teams accepting nonlinear progress, a clear contrast to traditional agile's ``done" increments after each cycle. For instance, the rollback mechanisms in \cite{A23-lei2020agile, A24-alnoukari2008applying} acknowledge ML's experimental nature but require cultural shifts in sprint planning. It is worth evaluating the hybrid approaches against \cite{A18-saltz2022achieving, A11-saltz2019ski} contributions that suggest decoupling ceremonies from sprints may mitigate misalignment between ML tasks and time-boxed iterations, leading to minor cultural shifts within the team.

Task estimation remains problematic due to the experimental nature of ML tasks, and the rigid time-boxing of sprints underscores a fundamental tension between agile's predictability and ML's exploratory demands. This echoes broader critiques about applying time-boxed iterations to research-intensive tasks \cite{A19-saltz2021identifying}. However, proposed adaptations like capability-based iterations \cite{A11-saltz2019ski,A18-saltz2022achieving} remain under-evaluated.

Additional challenges include a lack of standardized guidelines and insufficient cross-disciplinary training, which hampers effective communication and collaboration between data scientists and software engineers. The misalignment between business and technical objectives further complicates the development process, while ethical considerations, although recognized, are yet to be robustly operationalized within agile frameworks. 

These issues not only hinder the practical adoption of agile practices in ML projects but also highlight fertile ground for future research. First, there is a need to reconcile the inherently sequential and interdependent nature of machine learning workflows with the iterative cadence of agile methodologies. Another important direction is the development of generalizable agile guidelines and tools tailored for ML-enabled systems. Improving collaboration and alignment between ML teams and business stakeholders also remains an important concern. In addition, research could help to answer how to operationalize ethical considerations within agile ML development practices. Finally, there is an opportunity to advance methods for task prioritization and effort estimation to further support the effective integration of ML into agile development.

% Ethical considerations, though addressed through modified user story templates [A09], face implementation barriers. Only 3 papers integrated ethics into agile artifacts, and none provided evidence of longitudinal ethical impact assessments.

% The prevalence of separate ML and SW teams in approaches like Agile4MLS [A02] highlights unresolved collaboration challenges. While cross-functional team structures are advocated in agile literature, our findings indicate that specialized ML roles persist, risking communication silos. The rotating team member strategy [A02] offers a partial solution but requires validation across organizational contexts.

Although the authors proposed new agile approaches and recommendations for managing ML-enabled system projects, a critical gap lies in the empirical validation of the proposed approaches. While 17 of 27 papers incorporated some form of evaluation (Fig. \ref{evaluationperyear}), the predominance of case studies (nine papers) over controlled experiments (one paper) limits their generalizability. Furthermore, 34.6\% of studies lacked any empirical evaluation, mirroring trends observed in related domains where solution proposals outpace rigorous validation \cite{wieringa2006requirements}. This imbalance between proposed methodologies and empirical evidence presents both a challenge for practitioners and an opportunity for researchers to design studies that assess the feasibility and generalizability of these adaptations.

\section{Threats to validity}

In this section, we discuss potential threats to the validity of our SMS and report actions taken to mitigate these risks. In terms of \textit{Internal Validity}, the search string may not be appropriately designed to suit the SMS. As outlined in Section III, we built the search string using the PICO strategy, ensuring its direct alignment with our research objectives. To further ensure the robustness of our approach, we adopted a hybrid search strategy, integrating database search and snowballing techniques. This methodology has been shown to be effective for secondary studies, producing an unbiased and representative collection of papers \cite{wohlin2022successful}.

Regarding \textit{External Validity}, we have sought to strengthen the generalizability and replicability of our findings; we made an effort to document the steps undertaken and the results obtained. this information is accessible through our Zenodo repository \footnote{\url{https://doi.org/10.5281/zenodo.14105692}}. Despite this effort, there's the possibility we have missed studies. To mitigate this risk, we conducted the hybrid search strategy multiple times throughout 2024, increasing the likelihood of finding any missed studies. Given the repeated application of this search process, we are confident that our set of identified primary studies is comprehensive.

Finally, \textit{Reliability} concerns center on the potential bias in study selection. To mitigate this risk, all steps were conducted by more than one researcher, including study selection, data extraction, and coding (which was independently peer reviewed). Any discrepancies were resolved through discussion and consensus. Furthermore, the open science repository helps to make our research steps auditable.

\section{Concluding Remarks}

In this paper, we presented an SMS on agile management for ML-enabled systems development. We performed a hybrid search strategy that combines a Scopus database search with iterative backward and forward snowballing; we identified 27 studies addressing agile practices tailored to the unique challenges of ML development.

Our analysis categorized the literature agile management approaches, tailored practices, recommendations, challenges, and the types of empirical studies conducted. The findings reveal that while innovative adaptations, such as hybrid agile-lifecycle integrations and decoupled scrum ceremonies, are emerging to bridge the gap between traditional agile methods and ML workflows, significant challenges remain in areas like effort estimation and cross-disciplinary collaboration.

The main contributions of this study are twofold: first, providing a comprehensive mapping of the state of the art in agile management for ML-enabled systems, and second, identifying key research gaps that invite further investigation, the challenges related to effort estimation, sprint planning, process interdependencies, and ethical operationalization suggest that a one-size-fits-all approach remains elusive. We believe that these insights offer valuable guidance for practitioners and set the stage for future research aimed at empirically validating and refining agile practices to better support ML-enabled system development.

\bibliographystyle{IEEEtranS}
\bibliography{refs}

% Generated by IEEEtranS.bst, version: 1.14 (2015/08/26)
\begin{thebibliography}{10}
\providecommand{\url}[1]{#1}
\csname url@samestyle\endcsname
\providecommand{\newblock}{\relax}
\providecommand{\bibinfo}[2]{#2}
\providecommand{\BIBentrySTDinterwordspacing}{\spaceskip=0pt\relax}
\providecommand{\BIBentryALTinterwordstretchfactor}{4}
\providecommand{\BIBentryALTinterwordspacing}{\spaceskip=\fontdimen2\font plus
\BIBentryALTinterwordstretchfactor\fontdimen3\font minus \fontdimen4\font\relax}
\providecommand{\BIBforeignlanguage}[2]{{%
\expandafter\ifx\csname l@#1\endcsname\relax
\typeout{** WARNING: IEEEtranS.bst: No hyphenation pattern has been}%
\typeout{** loaded for the language `#1'. Using the pattern for}%
\typeout{** the default language instead.}%
\else
\language=\csname l@#1\endcsname
\fi
#2}}
\providecommand{\BIBdecl}{\relax}
\BIBdecl

\bibitem{ahmad2013kanbansms}
M.~O. Ahmad, J.~Markkula, and M.~Oivo, ``Kanban in software development: A systematic literature review,'' in \emph{2Euromicro Conference on Software Engineering and Advanced Applications}, 2013, pp. 9--16.

\bibitem{A24-alnoukari2008applying}
M.~Alnoukari, Z.~Alzoabi, and S.~Hanna, ``Applying adaptive software development (asd) agile modeling on predictive data mining applications: Asd-dm methodology,'' in \emph{International Symposium on Information Technology}, vol.~2, 2008, pp. 1--6.

\bibitem{alves2023practices}
I.~Alves, L.~A. Leite, P.~Meirelles, F.~Kon, and C.~S.~R. Aguiar, ``Practices for managing machine learning products: A multivocal literature review,'' \emph{IEEE Transactions on Engineering Management}, vol.~71, pp. 7425--7455, 2023.

\bibitem{A12-baijens2020applying}
J.~Baijens, R.~Helms, and D.~Iren, ``Applying scrum in data science projects,'' in \emph{IEEE Conference on Business Informatics (CBI)}, vol.~1, 2020, pp. 30--38.

\bibitem{A16-baijens2020data}
J.~Baijens, R.~Helms, and R.~Kusters, ``Data analytics project methodologies: Which one to choose?'' in \emph{International Conference on Big Data in Management}, 2020, pp. 41--47.

\bibitem{beck2001manifesto}
K.~Beck, M.~Beedle, A.~Van~Bennekum, A.~Cockburn, W.~Cunningham, M.~Fowler, J.~Grenning, J.~Highsmith, A.~Hunt, R.~Jeffries \emph{et~al.}, ``Manifesto for agile software development,'' 2001.

\bibitem{busquim2024interaction}
G.~Busquim, H.~Villamizar, M.~J. Lima, and M.~Kalinowski, ``On the interaction between software engineers and data scientists when building machine learning-enabled systems,'' in \emph{International Conference on Software Quality}, 2024, pp. 55--75.

\bibitem{cohn2004userstoriesapplied}
M.~Cohn, \emph{User stories applied: For agile software development}.\hskip 1em plus 0.5em minus 0.4em\relax Addison-Wesley Professional, 2004.

\bibitem{A26-10.1007/978-3-031-48550-3_6}
R.~Cordeiro, I.~Alves, S.~Alves, and A.~Goldman, ``Being agile in a data science project,'' in \emph{Agile Processes in Software Engineering and Extreme Programming -- Workshops}, P.~Kruchten and P.~Gregory, Eds., 2024.

\bibitem{cruzes2011recommended}
D.~S. Cruzes and T.~Dyba, ``Recommended steps for thematic synthesis in software engineering,'' in \emph{International Symposium on Empirical Software Engineering and Measurement}, 2011, pp. 275--284.

\bibitem{A15-dastgerdi2021appropriate}
A.~K. Dastgerdi and T.~J. Gandomani, ``On the appropriate methodologies for data science projects,'' in \emph{International Conference on Information Technology (ICIT)}, 2021, pp. 667--673.

\bibitem{A08-haakman2021ai}
M.~Haakman, L.~Cruz, H.~Huijgens, and A.~Van~Deursen, ``Ai lifecycle models need to be revised: An exploratory study in fintech,'' \emph{Empirical Software Engineering}, vol.~26, no.~5, p.~95, 2021.

\bibitem{A04-halme2021write}
E.~Halme, V.~Vakkuri, J.~Kultanen, M.~Jantunen, K.-K. Kemell, R.~Rousi, and P.~Abrahamsson, ``How to write ethical user stories? impacts of the eccola method,'' in \emph{International Conference on Agile Software Development}, 2021, pp. 36--52.

\bibitem{A07-jackson2019agile}
S.~Jackson, M.~Yaqub, and C.-X. Li, ``The agile deployment of machine learning models in healthcare,'' \emph{Frontiers in Big Data}, vol.~1, p.~7, 2019.

\bibitem{A09-kemell2022utilizing}
K.-K. Kemell, V.~Vakkuri, and E.~Halme, ``Utilizing user stories to bring ai ethics into practice in software engineering,'' in \emph{International Conference on Product-Focused Software Process Improvement}, 2022, pp. 553--558.

\bibitem{A03-kohl2021stamp}
P.~Kohl, O.~Schmidts, L.~Kl{\"o}ser, H.~Werth, B.~Kraft, and A.~Z{\"u}ndorf, ``Stamp 4 nlp--an agile framework for rapid quality-driven nlp applications development,'' in \emph{International Conference on the Quality of Information and Communications Technology}, 2021, pp. 156--166.

\bibitem{A14-kraut2022application}
N.~Kraut and F.~Transchel, ``On the application of scrum in data science projects,'' in \emph{International Conference on Big Data Analytics (ICBDA)}, 2022, pp. 1--9.

\bibitem{whatMakesAgileSoft.Agile}
M.~Kuhrmann, P.~Tell, R.~Hebig, and \textit{et al.}, ``What makes agile software development agile?'' \emph{IEEE Transactions on Software Engineering}, vol.~48, no.~9, pp. 3523--3539, 2022.

\bibitem{A21-lahiri2023evaluating}
S.~Lahiri and J.~Saltz, ``Evaluating data science project agility by exploring process frameworks used by data science teams,'' 2023.

\bibitem{A23-lei2020agile}
H.~Lei, R.~O’Connell, L.~Ehwerhemuepha, S.~Taraman, W.~Feaster, and A.~Chang, ``Agile clinical research: A data science approach to scrumban in clinical medicine,'' \emph{Intelligence-based medicine}, vol.~3, p. 100009, 2020.

\bibitem{A05-leijnen2020agile}
S.~Leijnen, H.~Aldewereld, R.~van Belkom, R.~Bijvank, and R.~Ossewaarde, ``An agile framework for trustworthy ai.'' in \emph{NeHuAI@ ECAI}, 2020, pp. 75--78.

\bibitem{leonardo2018pico}
R.~Leonardo, ``Pico: model for clinical questions,'' \emph{Evid Based Med Pract}, vol.~3, no. 115, p.~2, 2018.

\bibitem{mitchell1997machine}
\BIBentryALTinterwordspacing
T.~Mitchell, \emph{Machine Learning}, ser. McGraw-Hill International Editions.\hskip 1em plus 0.5em minus 0.4em\relax McGraw-Hill, 1997. [Online]. Available: \url{https://books.google.com.br/books?id=EoYBngEACAAJ}
\BIBentrySTDinterwordspacing

\bibitem{nahar2023meta-summary}
N.~Nahar, H.~Zhang, G.~Lewis, S.~Zhou, and C.~K{\"a}stner, ``A meta-summary of challenges in building products with ml components--collecting experiences from 4758+ practitioners,'' in \emph{International Conference on AI Engineering--Software Engineering for AI (CAIN)}, 2023, pp. 171--183.

\bibitem{petersen2015guidelines}
K.~Petersen, S.~Vakkalanka, and L.~Kuzniarz, ``Guidelines for conducting systematic mapping studies in software engineering: An update,'' \emph{Information and Software Technology}, vol.~64, pp. 1--18, 2015.

\bibitem{A13-qadadeh2020improved}
W.~Qadadeh and S.~Abdallah, ``An improved agile framework for implementing data science initiatives in the government,'' in \emph{International Conference on Information and Computer Technologies (ICICT)}, 2020, pp. 24--30.

\bibitem{A10-saltz2017comparing}
J.~Saltz, K.~Crowston \emph{et~al.}, ``Comparing data science project management methodologies via a controlled experiment,'' 2017.

\bibitem{A18-saltz2022achieving}
J.~Saltz, A.~Sutherland, and N.~Hotz, ``Achieving lean data science agility via data driven scrum,'' 2022.

\bibitem{A11-saltz2019ski}
J.~Saltz and A.~Suthrland, ``Ski: An agile framework for data science,'' in \emph{IEEE International Conference on Big Data (Big Data)}, 2019, pp. 3468--3476.

\bibitem{A20saltz2020identifying}
J.~S. Saltz and N.~Hotz, ``Identifying the most common frameworks data science teams use to structure and coordinate their projects,'' in \emph{International Conference on Big Data (Big Data)}, 2020, pp. 2038--2042.

\bibitem{A17-saltz2019achieving}
J.~S. Saltz and I.~Shamshurin, ``Achieving agile big data science: the evolution of a team’s agile process methodology,'' in \emph{IEEE International Conference on Big Data (Big Data)}, 2019, pp. 3477--3485.

\bibitem{A19-saltz2021identifying}
J.~S. Saltz, A.~Sutherland, and T.~Jombart, ``Identifying and addressing 6 key questions when using data driven scrum,'' in \emph{International Conference on Big Data (Big Data)}, 2021, pp. 2345--2352.

\bibitem{A22-schmidt2018synthesizing}
C.~Schmidt and W.~N. Sun, ``Synthesizing agile and knowledge discovery: case study results,'' \emph{Journal of Computer Information Systems}, vol.~58, no.~2, pp. 142--150, 2018.

\bibitem{schwaber1997scrum}
K.~Schwaber, ``Scrum development process,'' in \emph{Business Object Design and Implementation: OOPSLA’95 Workshop Proceedings}, 1997, pp. 117--134.

\bibitem{A06-singla2018analysis}
K.~Singla, J.~Bose, and C.~Naik, ``Analysis of software engineering for agile machine learning projects,'' in \emph{IEEE India Council International Conference (INDICON)}, 2018, pp. 1--5.

\bibitem{A25-uysal2022machine}
M.~P. Uysal, ``Machine learning and data science project management from an agile perspective: Methods and challenges,'' in \emph{Contemporary challenges for agile project management}.\hskip 1em plus 0.5em minus 0.4em\relax IGI Global, 2022, pp. 73--88.

\bibitem{A27uysal2023toward}
------, ``Toward a method engineering framework for project management and machine learning,'' in \emph{Annual Computers, Software, and Applications Conference (COMPSAC)}, 2023, pp. 1186--1190.

\bibitem{A02-vaidhyanathan2022agile4mls}
K.~Vaidhyanathan, A.~Chandran, H.~Muccini, and R.~Roy, ``Agile4mls—leveraging agile practices for developing machine learning-enabled systems: An industrial experience,'' \emph{IEEE Software}, vol.~39, no.~6, pp. 43--50, 2022.

\bibitem{A01-vial2023managing}
G.~Vial, A.-F. Cameron, T.~Giannelia, and J.~Jiang, ``Managing artificial intelligence projects: Key insights from an ai consulting firm,'' \emph{Information Systems Journal}, vol.~33, no.~3, pp. 669--691, 2023.

\bibitem{wieringa2006requirements}
R.~Wieringa, N.~Maiden, N.~Mead, and C.~Rolland, ``Requirements engineering paper classification and evaluation criteria: a proposal and a discussion,'' \emph{Requirements Engineering}, vol.~11, pp. 102--107, 2006.

\bibitem{wohlin2022successful}
C.~Wohlin, M.~Kalinowski, K.~R. Felizardo, and E.~Mendes, ``Successful combination of database search and snowballing for identification of primary studies in systematic literature studies,'' \emph{Information and Software Technology}, vol. 147, p. 106908, 2022.

\bibitem{wohlin2024experimentation}
C.~Wohlin, P.~Runeson, M.~H{\"o}st, M.~C. Ohlsson, B.~Regnell, and A.~Wessl{\'e}n, \emph{Experimentation in Software Engineering}.\hskip 1em plus 0.5em minus 0.4em\relax Springer, 2024.

\end{thebibliography}

\end{document}